\documentclass[pra,aps,twocolumn]{revtex4}
\usepackage{graphicx}
\begin{document}
\title{Free-carrier driven spatio-temporal dynamics in amplifying silicon waveguides}
\author{Samudra Roy$^{1,2}$, Andrea Marini$^1$ and Fabio Biancalana$^{1,3}$}
\affiliation{$^1$Max Planck Institute for the Science of Light, G$\ddot{u}$nther-Scharowsky-Stra\ss e 1, 91058 Erlangen, Germany}
\affiliation{$^2$Department of Physics and Meteorology, Indian Institute of Technology, Kharagpur-721302, India}
\affiliation{$^3$School of Engineering \& Physical Sciences, Heriot-Watt University, EH14 4AS Edinburgh, United Kingdom}
\date{\today}
\begin{abstract}
We theoretically investigate the free-carrier induced spatio-temporal dynamics of continuous waves in silicon
waveguides embedded in an amplifying medium. Optical propagation is governed by a cubic
Ginzburg-Landau equation coupled with an ordinary differential equation accounting for the free-carrier
dynamics. We find that, owing to free-carrier dispersion, countinuous waves are modulationally unstable 
in both anomalous and normal dispersion regimes and chaotically generate unstable accelerating pulses. 
\end{abstract}
\maketitle

\section{Introduction}

Silicon photonics is a well established area of research aiming at exploiting silicon (Si)
as a photonic component for the engineering of integrated optoelectronic devices. The extraordinary 
optical properties of Si open up possibilities for novel miniaturized applications,
ranging from optical interconnection to bio-sensing \cite{jalali,soref}. In the mid-infrared,
Si has a high refractive index ($n \simeq 3.5$) and negligible linear extinction. However, in the range 
1 $\mu$m $< \lambda_0 < $ 2.2 $\mu$m, two photon absorption (2PA) is relevant and is responsible for 
high nonlinear extinction \cite{soref2}. Owing to the high refractive index, light can be tightly localized 
in subwavelength Si-based waveguides \cite{Cheben}, which tremendously enhance nonlinear processes \cite{leuthold}, 
including 2PA that damps optical propagation and limits the efficiency of Si-based photonic components \cite{yin}. 
As a consequence of 2PA, pairs of photons with total energy greater than the Si band-gap ($E_g\approx1.12$ eV) are 
absorbed and electrons are excited to the conduction band modifying the Si optical response \cite{lin2}. In this 
interesting and rather unexplored operating regime, free-carriers (FCs) directly interact with the optical field 
and introduce novel nonlinear effects \cite{husko2}. A quite similar scenario happens in gas-filled hollow 
core photonic crystal fibers (HCPCFs) in the ionization regime, where accelerating solitons \cite{saleh} 
and universal modulational instability \cite{saleh1} have been recently observed. In this case, the free 
plasma generated through ionization is responsible for an intense {\it blueshift} of several hundreds of 
nanometers of the optical pulse \cite{holzer}. This tremendous dynamics occurs in a much smaller scale 
(blueshift of few nanometers) in Si-based waveguides \cite{husko2} because of the intimate presence of 2PA, 
which is responsible for both the creation of FCs and for damping. In principle, losses can be 
reduced in hybrid slot waveguides \cite{koos}, but in this case also the extraordinary effects
ensuing from FC dynamics are reduced accordingly. 

An alternative strategy for overcoming losses consists of embedding Si waveguides in gaining media. 
In this context, amplification schemes based on III-V semiconductors \cite{fang}, rare-earth-ion-doped 
dielectric thin films \cite{worhoff} and erbium-doped waveguides \cite{agazzi} have been proposed and 
practically realized. The paradigm model for describing optical propagation in amplifying waveguides is 
represented by the cubic Ginzburg-Landau (GL) equation, which governs a wide range of dissipative phenomena
\cite{akhmedievbook,aranson}. In general, GL systems are rich in nature and exhibit some peculiar features, 
e.g., chaos and pattern formation \cite{kishiba,saarloos}. Conversely to the case of 
conventional Kerr solitons arising from the balance between nonlinearity and dispersion \cite{agrawalbook}, 
localized stationary solutions of GL systems - namely {\it dissipative solitons} - result from the exact
compensation of gain and loss \cite{pereira}. In a recent work, we investigated the FC-induced
dynamics of dissipative solitons in Si-based amplifying waveguides, demonstrating the self-frequency 
soliton blueshift \cite{roy}.

In this paper, we theoretically investigate the propagation of continuous waves (CWs) in a 
silicon-on-insulator (SOI) waveguide embedded in erbium-doped amorphous aluminium oxide (Al$_2$O$_3$:Er$^+$). 
Owing to the externally pumped active inclusions, small optical waves are exponentially amplified and 
instability develops until the nonlinear gain saturation comes into play counterbalancing the linear 
amplification. Taking full account of FC generation and recombination, we calculate the stationary 
nonlinear CWs of the system and we investigate their stability. We find that, analogously to 
gas-filled HCPCFs \cite{saleh1}, stationary CWs are universally unstable in both normal and anomalous 
dispersion regimes. However, due to the inherent non-conservative nature of our system, modulational 
instability (MI) does not generate a shower of solitons as in Ref. \cite{saleh1}, but an {\it accelerating 
chaotic state}. This scenario ensues from the presence of unstable dissipative solitons, which constitute 
the strange attractor of the system \cite{kishiba}. Every sub-pulse generated through MI is accelerated by
the FC dispersion and experiences self-frequency blue-shift. In turn, the overall dynamics accelerates in 
the temporal domain and blueshifts in the frequency domain. The paper is organized as follows. In section II 
we describe the geometry of the system, introducing the governing equations and calculating the nonlinear 
stationary CW modes. In section III we analytically and numerically study MI of CWs. Analytical calculations 
predict universal MI and accelerating chaos, which is confirmed by numerical simulations.

\section{Model}

We consider a SOI waveguide with lateral dimensions $h = w = 525$ nm surrounded by Al$_2$O$_3$:Er$^+$, which 
gain bandwidth is of the order of $100$ nm around the carrier wavelength $\lambda \simeq 1540 $ nm. In 
principle, other gain schemes involving the use of semiconductor active materials can be considered and the 
gain bandwidth can be increased accordingly \cite{gmachl}. Without any loss of generality, we assume that 
the SOI waveguide is fabricated along the $[\bar{1}10]$ direction and on the $[110]\times[001]$ surface. In
this case, stimulated Raman scattering (SRS) does not occur for quasi-TM modes \cite{lin2}. Initially 
neglecting gain of the external medium, Si nonlinearity and FC generation, we have numerically calculated the 
linear quasi-TM mode ${\bf e}({\bf r}_{\bot})$, its dispersion $\beta_0(\omega_0)$ and the effective area by 
using COMSOL. At $\lambda_0 = 1550 $ nm, we have then found that the second order group velocity dispersion 
(GVD) coefficient is $\beta_2 \simeq -2$ ps$^2$/m and the effective area is $A_{\rm eff} \simeq 0.145$ $\mu$m$^2$.
Nonlinear pulse propagation can be modeled within the slowly varying envelope approximation (SVEA) by
taking the Ansatz for the electric field
${\bf E}({\bf r},t) = {\rm Re} \left [A(z,t){\bf e}({\bf r}_{\bot}) e^{i\beta_0 z - i\omega_0 t}\right]$,
where $A(z,t)$ is the pulse envelope, ${\bf e}({\bf r}_{\bot})$ is the linear mode profile,
$\omega_0 = 2\pi c/\lambda_0$ is the carrier angular frequency, $c$ is the speed of light in vacuum and
$\beta_0$ is the linear propagation constant. We approximate the external gaining medium as a
two-level system, which is characterized by a Lorentzian spectral distribution of gain around the carrier
angular frequency $\omega_0$: $g(\Delta\omega) = g_0/(1+\Delta\omega^2 T_2^2)$, where $g_0$ is the
dimensionless gain peak, $T_2 \simeq 40$ fs is the dephasing time of Al$_2$O$_3$:Er$^+$ and
$\Delta\omega = \omega - \omega_0$ \cite{agrawal}. For small detuning $\Delta\omega<<\omega_0$, the
spectral gain distribution can be approximated by $g(\Delta\omega) \simeq g_0 - g_2 \Delta\omega^2T_2^2$.
FCs affect the optical propagation by means of two mechanisms: FC-induced dispersion (FCD) and absorption (FCA). 
These processes are intrinsically nonlinear and the optical propagation is governed by the 
coupled equations
\begin{eqnarray}
&& i \partial_{\xi} u - \frac{s}{2}\partial_{\tau}^2u + i \alpha u + ( 1 + i K ) |u|^2 u \label{CGLE} \\
&& - i ( g_0 + g_2 \partial_{\tau}^2 )u  + ( i/2 - \mu ) \phi_c u = 0 , \nonumber \\
&& \frac{d \phi_c}{d \tau} = \theta_c |u|^4 - \frac{\phi_c}{\tau_c }. \label{FCE}
\end{eqnarray}
Time duration and propagation length are normalized to the initial pulse width $t_0=100$ fs
and dispersion length $L_{\rm D}=t_0^2/|\beta_2|$. The envelope amplitude ($A$) is normalized to
$u = A/\sqrt{P_0} $, where
$P_0 = \lambda_0 A_{\rm eff}/(2\pi n_2L_{\rm D})$ and $n_2\simeq( 4 \pm 1.5 ) \times 10^{-18} $m$^2$/W is
the Kerr nonlinear coefficient of bulk silicon \cite{Dinu}. In Eq. (\ref{CGLE}), the parameter $s = \pm 1$ 
represents the sign of the GVD ($s=+1$ normal dispersion and $s=-1$ anomalous dispersion), while the parameter
$K=\beta_{\rm TPA}\lambda_0/(4\pi n_2)$ is the 2PA coefficient, where $\beta_{\rm TPA} \simeq 8 \times 10^{-12}$m/W  
is the bulk 2PA constant \cite{Dinu}. The linear loss coefficient ($\alpha_l \simeq 0.2$ dB/cm) is renormalized to 
the dispersion length ($\alpha = \alpha_l L_{\rm D}$) and can be neglected at $\lambda\simeq1.55\mu$m, where 2PA dominates. 
The FC density ($N_c$) is normalized to $\phi_c = \sigma N_c L_{\rm D}$, where 
$\sigma \simeq 1.45 \times 10^{-21} $m$^2$ \cite{Rong}. The parameter
$\theta_c=\beta_{\rm TPA}|\beta_{2}|\sigma\lambda_0^3/(16\pi^3c\hbar t_0n_2^2)$ accounts for FCA,
while $\mu = 2\pi k_c/(\sigma\lambda_0)$ accounts for FCD, where
$k_c\simeq1.35\times10^{-27} $m$^3$ \cite{Dinu,lin}. The characteristic FC recombination time ($t_c=1$ ns)
is also normalized to the initial pulse width: $\tau_c = t_c/t_0 = 1\times10^4$.

\begin{figure}
\centering
\begin{center}
\includegraphics[width=0.45\textwidth] {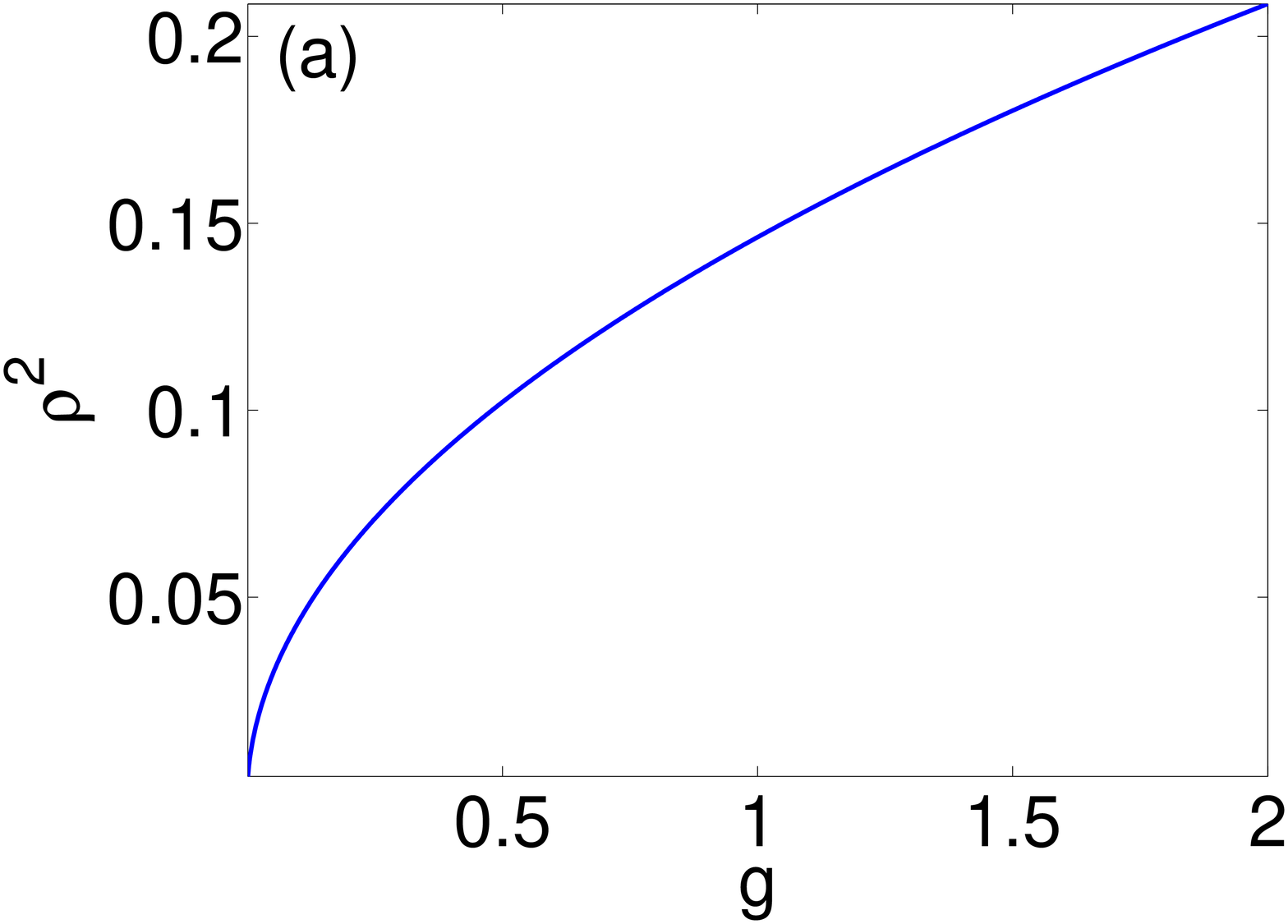}
\includegraphics[width=0.45\textwidth] {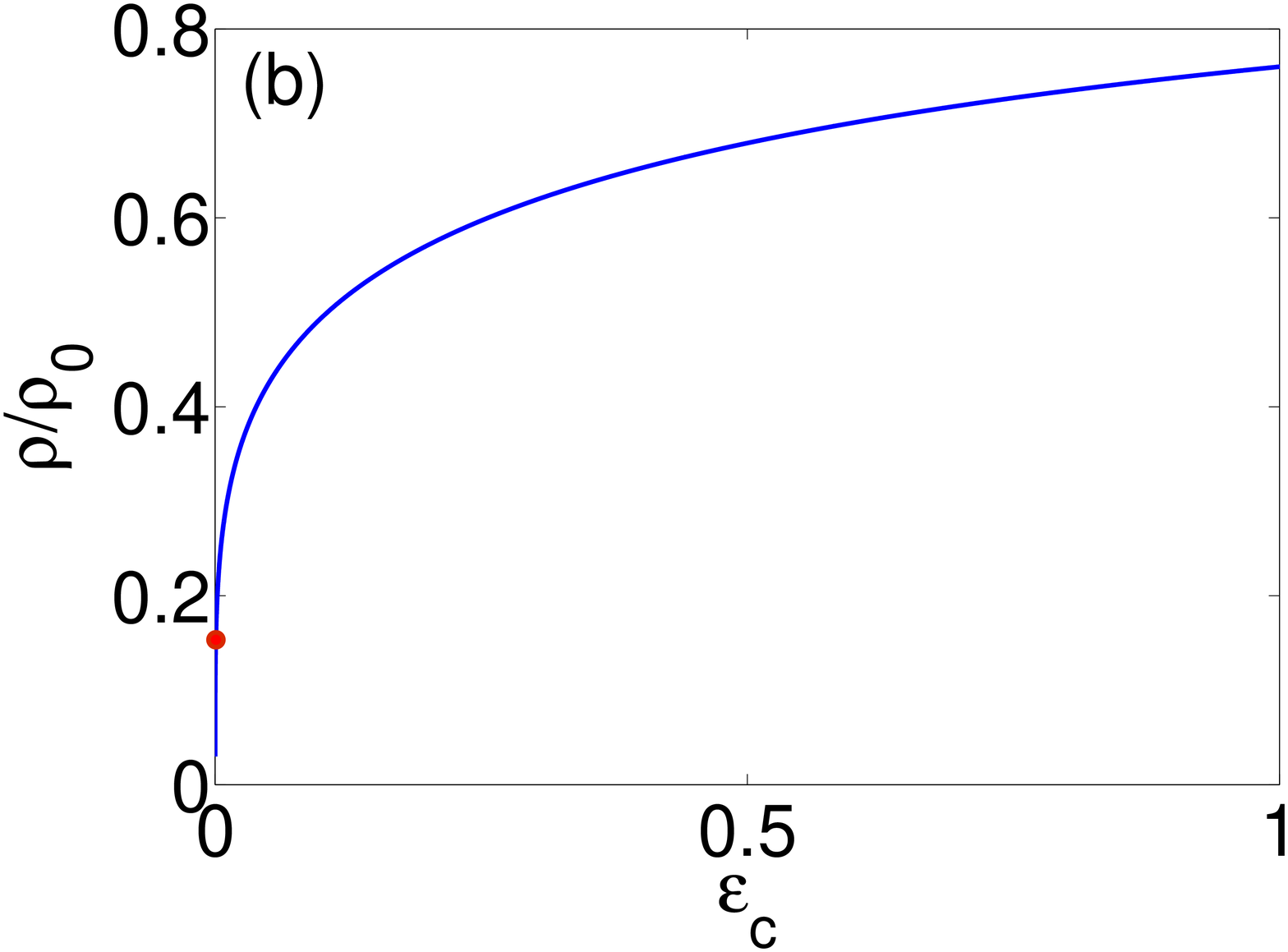}
\caption{(a) Squared amplitude $\rho^2$ of the stationary nonlinear CW mode as a function of the normalized
gain parameter of the external medium $g$. (b) Dependence of the rescaled mode amplitude $\rho/\rho_0$ on the
FC parameter $\epsilon_c$ for $g=1,g_2=0.16$.}
\label{Fig1}
\end{center}
\end{figure}

The stationary CW nonlinear modes of Eqs. (\ref{CGLE},\ref{FCE}) are found by setting the Ansatz
\begin{eqnarray}
&& u = \rho e^{i\eta\xi}, \\
&& \phi_c = \phi_0,
\end{eqnarray}
where $\rho,\eta$ are the mode amplitude and propagation constant, and $\phi_0=\theta_c\tau_c\rho^4$ is
the stationary number of FCs: $d\phi_c/d \tau = 0$. Inserting the Ansatz into 
Eqs. (\ref{CGLE},\ref{FCE}) one gets
\begin{eqnarray}
\rho^2 & = & - \epsilon_c + \sqrt{ \epsilon_c^2 + 2 \rho_0^2 \epsilon_c }, \label{rho} \\
\eta   & = &   \rho^2     + 2 \mu K( \rho^2 - \rho_0^2 ), \label{eta}
\end{eqnarray}
where $\epsilon_c=K/(\theta_c\tau_c)$ and $\rho_0^2=g/K$. In absence of FCs ($\epsilon_c\rightarrow\infty$),
the mode amplitude $\rho$ converges to $\rho_0$, which coincides with the CW mode amplitude of the uncoupled 
cubic GL equation. Note that the solution above does not constitute a family, but an isolated fixed point, 
as it generally occurs in dissipative systems \cite{akhmedievbook}. In Fig. \ref{Fig1}a, the stationary 
nonlinear CW mode squared amplitude ($\rho^2$) is plotted against the gain parameter of the external medium ($g$). 
In Fig. \ref{Fig1}b, the mode amplitude rescaled to the GL amplitude ($\rho/\rho_0$) is plotted against the 
FC parameter $\epsilon_c$. Note that $\rho\rightarrow\rho_0$ in the limit where FCs are not excited 
($\epsilon_c\rightarrow\infty$). The red dot in the figure indicates the exact value of $\rho/\rho_0$ for 
$g=1,g_2=0.16$ and realistic parameters of Si.

\section{Modulational Instability}

Modulational instability (MI) of the nonlinear CW stationary mode given in Eqs. (\ref{rho},\ref{eta}) can be
determined by perturbing it with small time- and space-dependent waves:
\begin{eqnarray}
&& u = ( \rho + a ) e^{i \eta \xi}, \\
&& \phi_c = \phi_0 + b,
\end{eqnarray}
where $a,b$ are perturbations of the optical field and of the FC distribution.
Plugging the Ansatz above in Eqs. (\ref{CGLE},\ref{FCE}) and retaining only linear terms one achieves the following 
set of coupled equations for $a,b$:
\begin{eqnarray}
&& i \partial_{\xi} a = ( \eta + i g ) a + \left( \frac{s}{2} + i g \right) \partial_{\tau}^2a + \label{aE} \\
&& - ( 1 + i K ) \rho^2(2a+a^*) - \left( \frac{ i } {2} - \mu \right) ( \phi_0 a + \rho b) , \nonumber \\
&& \frac{ d b } { d \tau} = 2 \theta_c \rho^3( a + a^* ) - \frac {b } { \tau_c } . \label{bE}
\end{eqnarray}

\begin{figure}
\centering
\begin{center}
\includegraphics[width=0.45\textwidth]{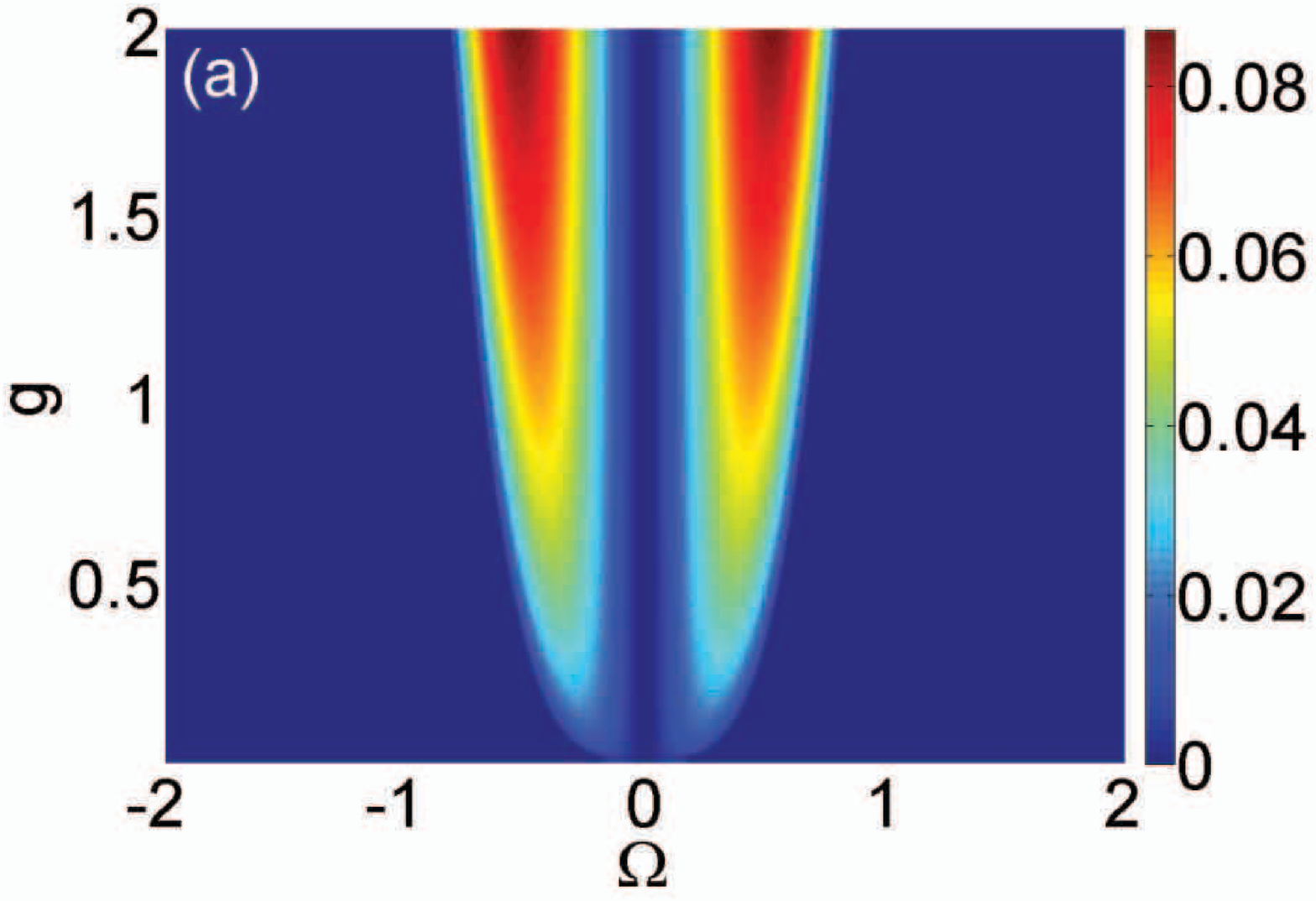}
\includegraphics[width=0.45\textwidth]{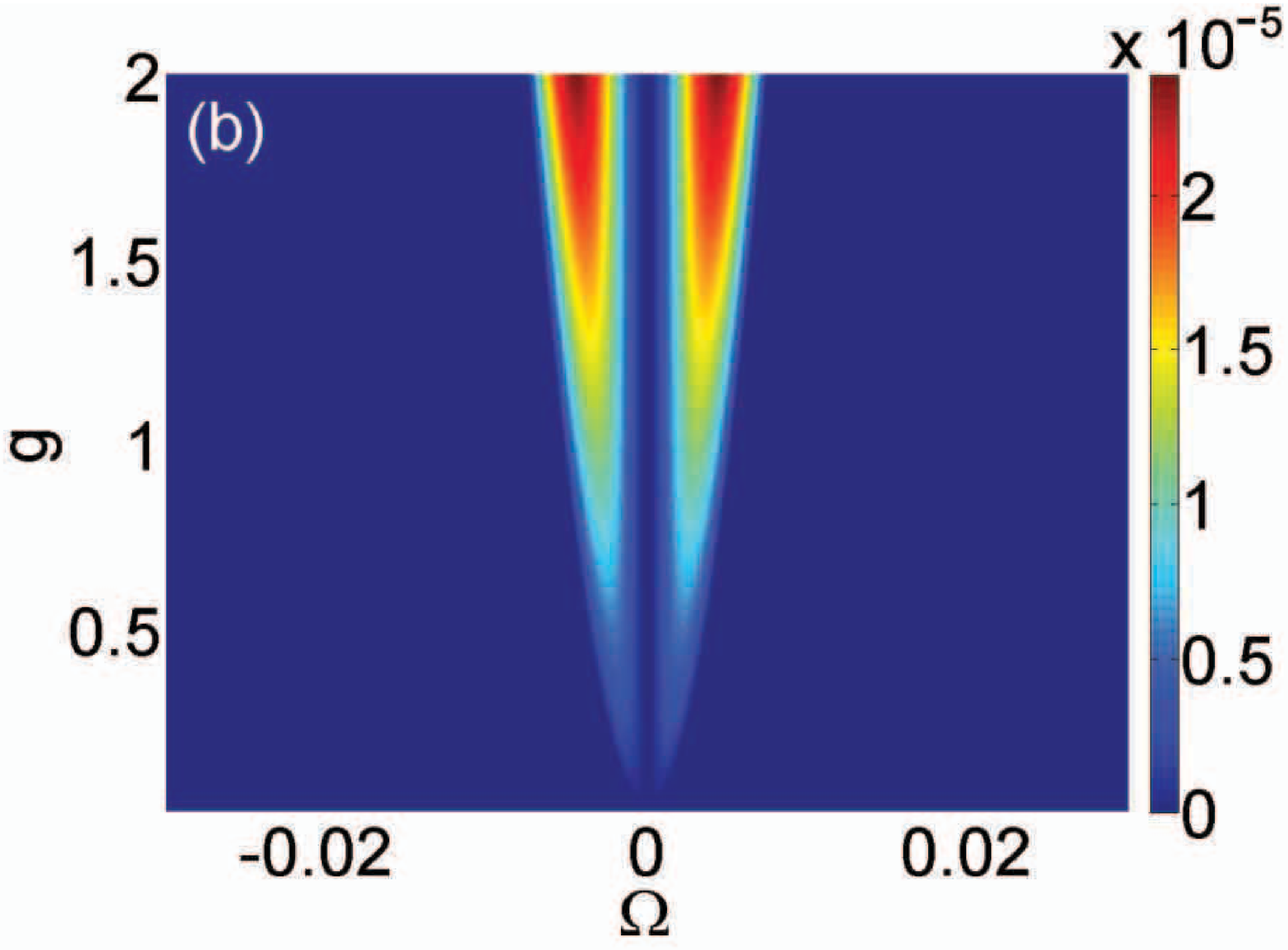}
\caption{Contour plots for the MI growth rate $h$ as a function of normalized gain $g$ and angular frequency $\Omega$
for (a) anomalous ($s=-1$) and (b) normal ($s=+1$) dispersion regimes. The second order GVD coefficient is $|\beta_2|=2$ ps$^2$/m, the 
gain dispersion is $g_2=0.16$, and the 2PA coefficient is $K \simeq 0.4$.}
\label{Fig2}
\end{center}
\end{figure}

We assume that $a = a_1 e^\vartheta + a_2^* e^{\vartheta^*}$, $b= b_0 e^{\vartheta} + b_0^* e^{\vartheta^*}$,
where $\vartheta = h \xi + i \Omega \tau$, $h$ is the growth rate of the small periodic perturbations
and $\Omega$ is their angular frequency. Note that $\phi_c$ is real and positive, since it 
represents the number of FCs generated via 2PA. Inserting the expressions above for $a,b$
in Eqs. (\ref{aE},\ref{bE}) we achieve a set of algebraic equations for the optical field perturbations $a_1,a_2$:
\begin{eqnarray}
\left[ \begin{array}{cc} i h + F + TL & P + T L \\ P^* + T^* L & - i h + F^* + T^* L \end{array} \right]\left[ \begin{array}{c} a_{1} \\ a_{2} \end{array} \right] = 0, \label{mat}
\end{eqnarray}
where $P = ( 1 + i K )  \rho^2$, $T = 2 \rho^4 \theta_c ( i/2 - \mu )$, and
\begin{eqnarray}
&& L = \frac{\tau_c^{-1} - i \Omega}{\tau_c^{-2} + \Omega^2}, \\
&& F = \left( \frac{s}{2} + i g \right) \Omega^2 + 2 P - \eta - i g + \left( \frac{i}{2} - \mu \right) \phi_0.
\end{eqnarray}
Non-trivial solutions can be found by setting the determinant of the coefficient matrix to zero, achieving the 
instability eigenvalues
\begin{eqnarray}
h_{1,2} = - A_0 (\Omega) \pm \sqrt{ A_0^2 (\Omega) - B_0 (\Omega) } \label{hE},
\end{eqnarray}
where
\begin{eqnarray}
A_0 & = & K \rho^2 + g_2 \Omega^2 + L \theta_c \rho^4 , \\
B_0 & = & \left( \frac{1}{4} + g_2^2 \right) \Omega^4 + ( 2 g_2 K + s ) \rho^2 \Omega^2 + \nonumber \\
    &   & + 2 L \theta_c ( g_2 - s \mu ) \rho^4 \Omega^2.
\end{eqnarray}

\begin{figure}
\centering
\begin{center}
\includegraphics[width=0.45\textwidth]{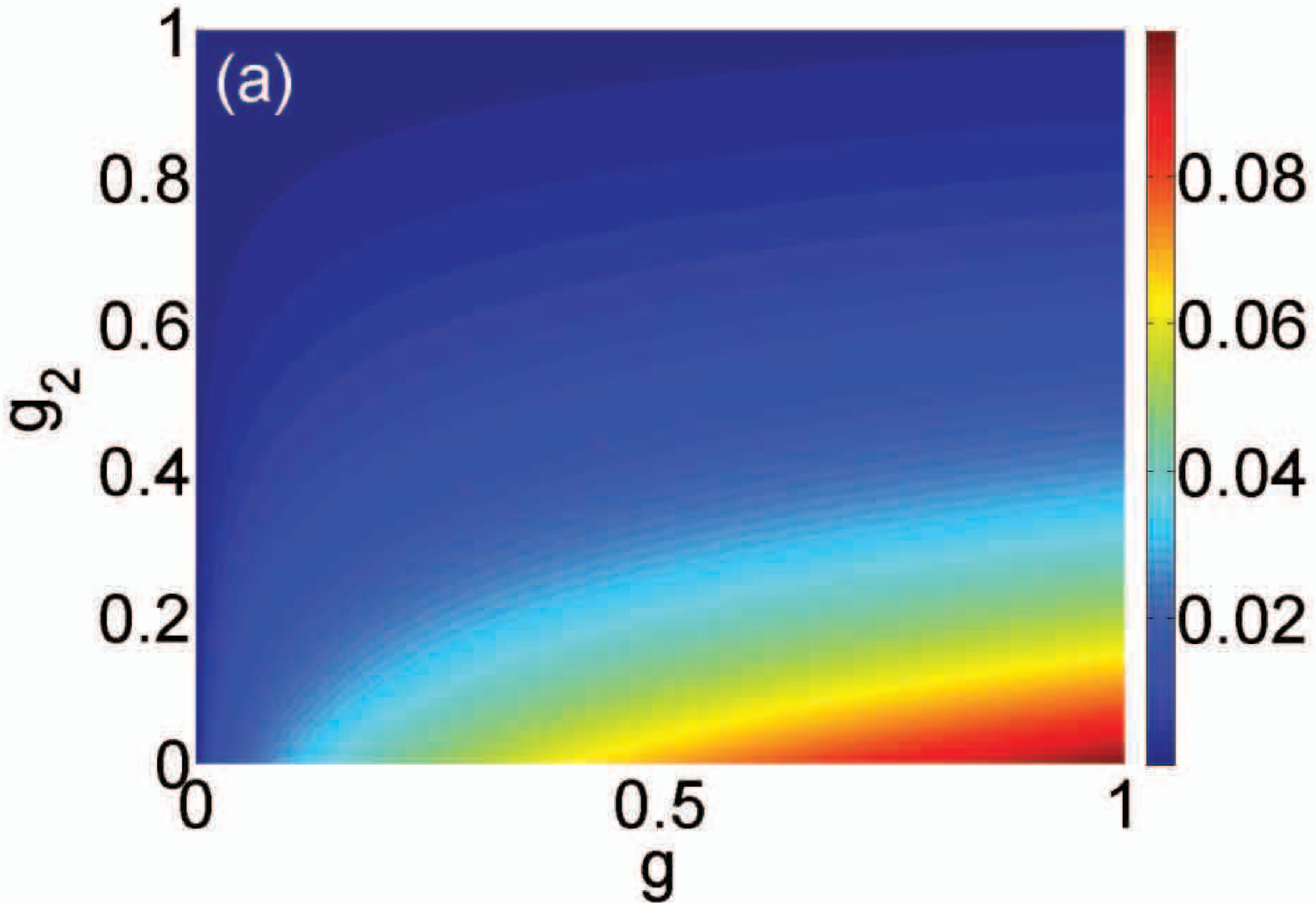}
\includegraphics[width=0.45\textwidth]{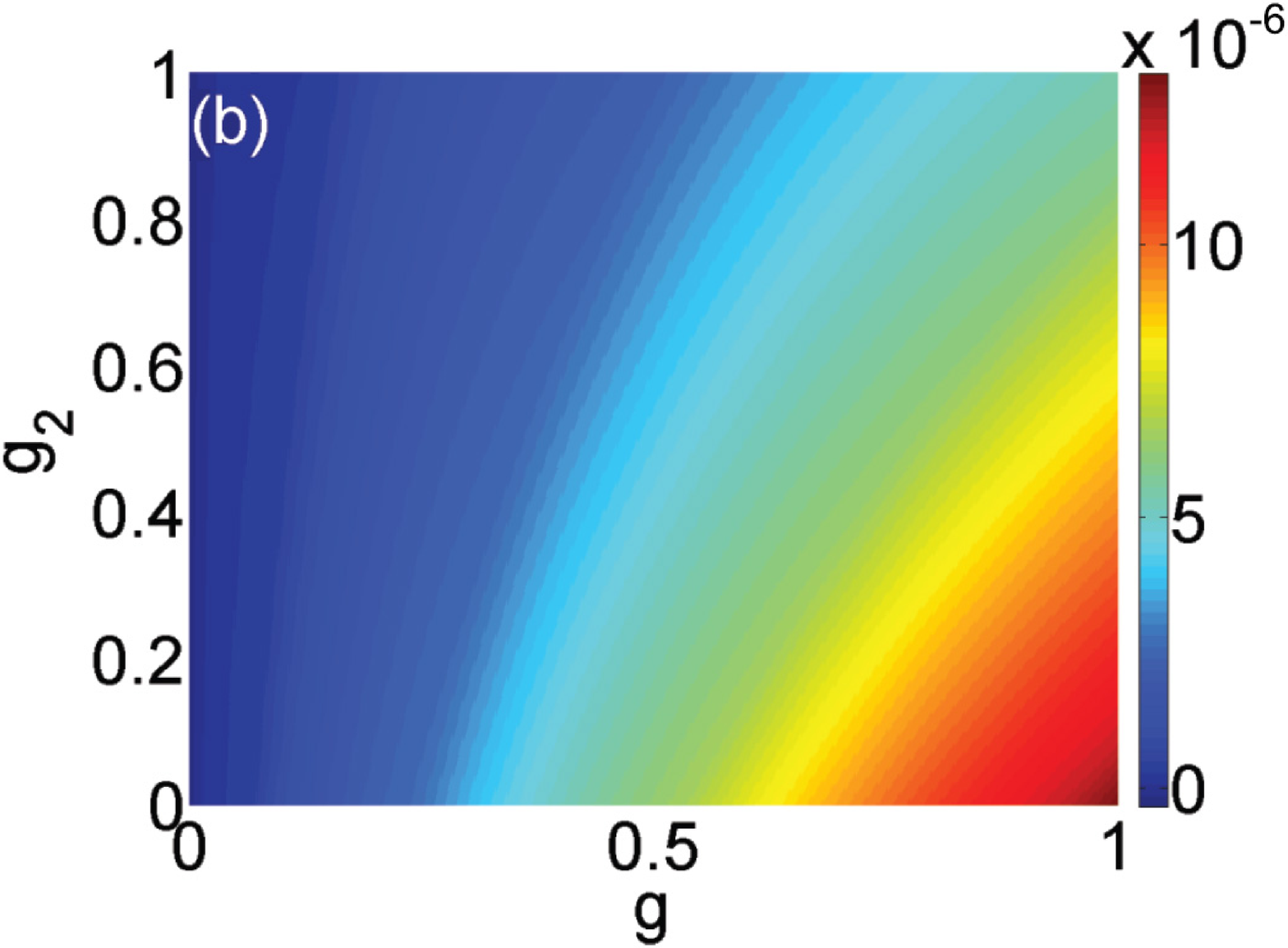}
\caption{Contour plot of the maximum instability eigenvalue ${\rm max}_{\Omega}h$ as a function of $g,g_2$ in
(a) anomalous ($s=-1$) and (b) normal ($s=+1$) dispersion regimes. The second order GVD coefficient is 
$|\beta_2|=2$ ps$^2$/m and the 2PA coefficient is $K \simeq 0.4$.}
\label{Fig3}
\end{center}
\end{figure}

If the real part of $h_1$ or $h_2$ is positive, small perturbations are amplified during propagation giving rise 
to instability. In Figs. \ref{Fig2}a, \ref{Fig2}b, we plot the instability parameter $h={\rm max} [{\rm Re} h_{1,2}]$ as 
a function of the normalized gain $g$ and angular frequency $\Omega$ for (a) anomalous (AD) and (b) normal (ND) 
dispersion. The interplay between nonlinear and dispersive
effects leads to instability across the central frequency by creating sidebands. Small
perturbations with angular frequencies falling within these spectral regions are amplified and
eventually the nonlinear stationary CW mode breaks up into a train of pulses. Owing to the refractive index change 
induced by FCs, these pulses are accelerated and we found that the instability eigenvectors are asymmetrically peaked 
on a blue-shifted frequency. Remarkably, FC-induced instability can be observed in ND regime, analogously to universal 
plasma-induced instability observed in gas-filled hollow core photonic crystal fibers \cite{saleh1}. In Fig. \ref{Fig3} 
we have plotted the maximum instability eigenvalue ${\rm max}_{\Omega}h$ as a function of $g,g_2$ for (a) anomalous and 
(b) normal dispersion. Note that in both regimes, the instability parameter increases with gain and is reduced 
by gain dispersion. While in AD generation of FCs counteracts instability, in ND it is the sole crucial term 
responsible for its occurrence. Note that also the vacuum background is unstable in supercritical conditions ($g_0>0$). 

\begin{figure}
\centering
\begin{center}
\includegraphics[width=0.45\textwidth]{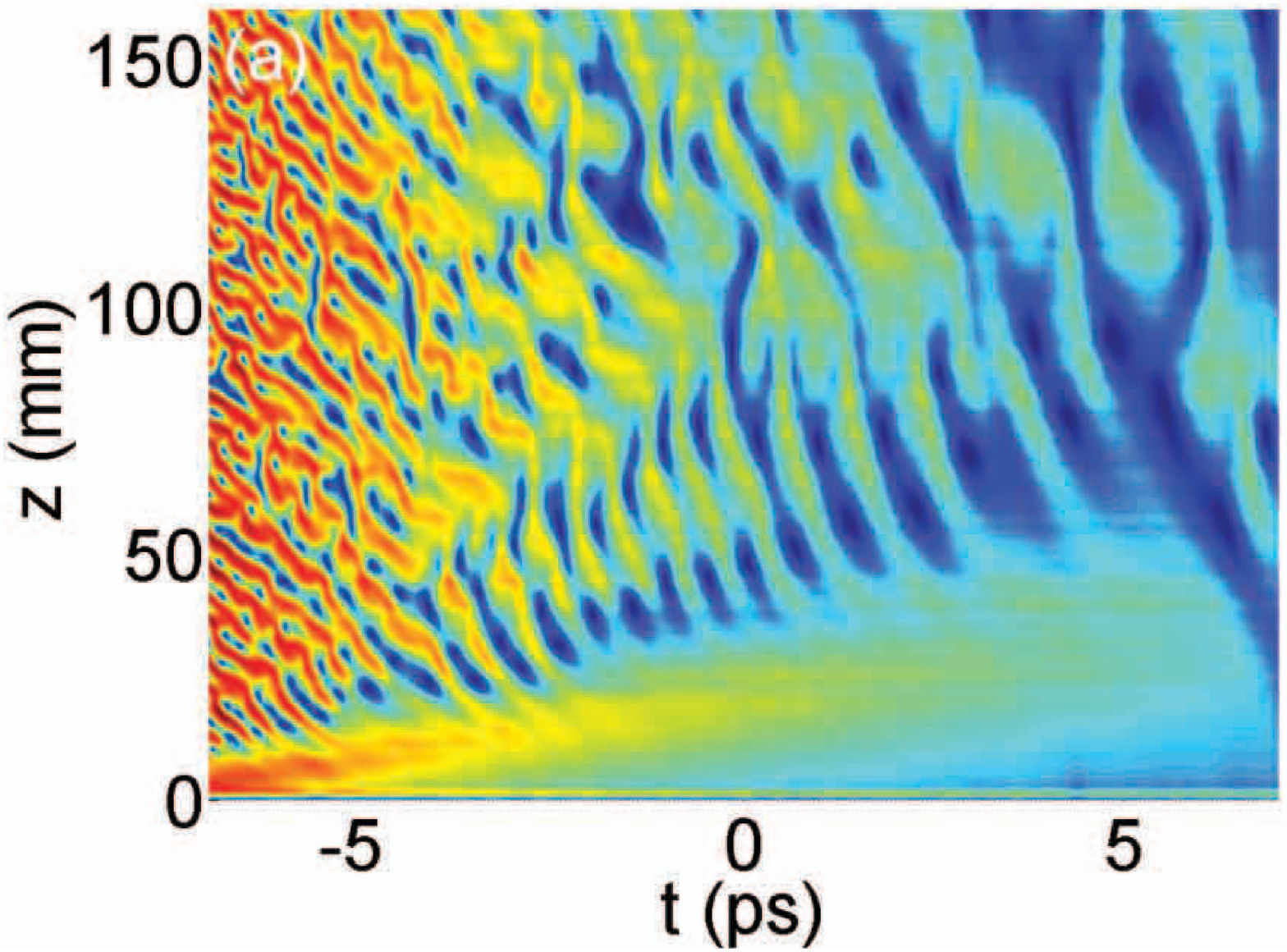}
\includegraphics[width=0.45\textwidth]{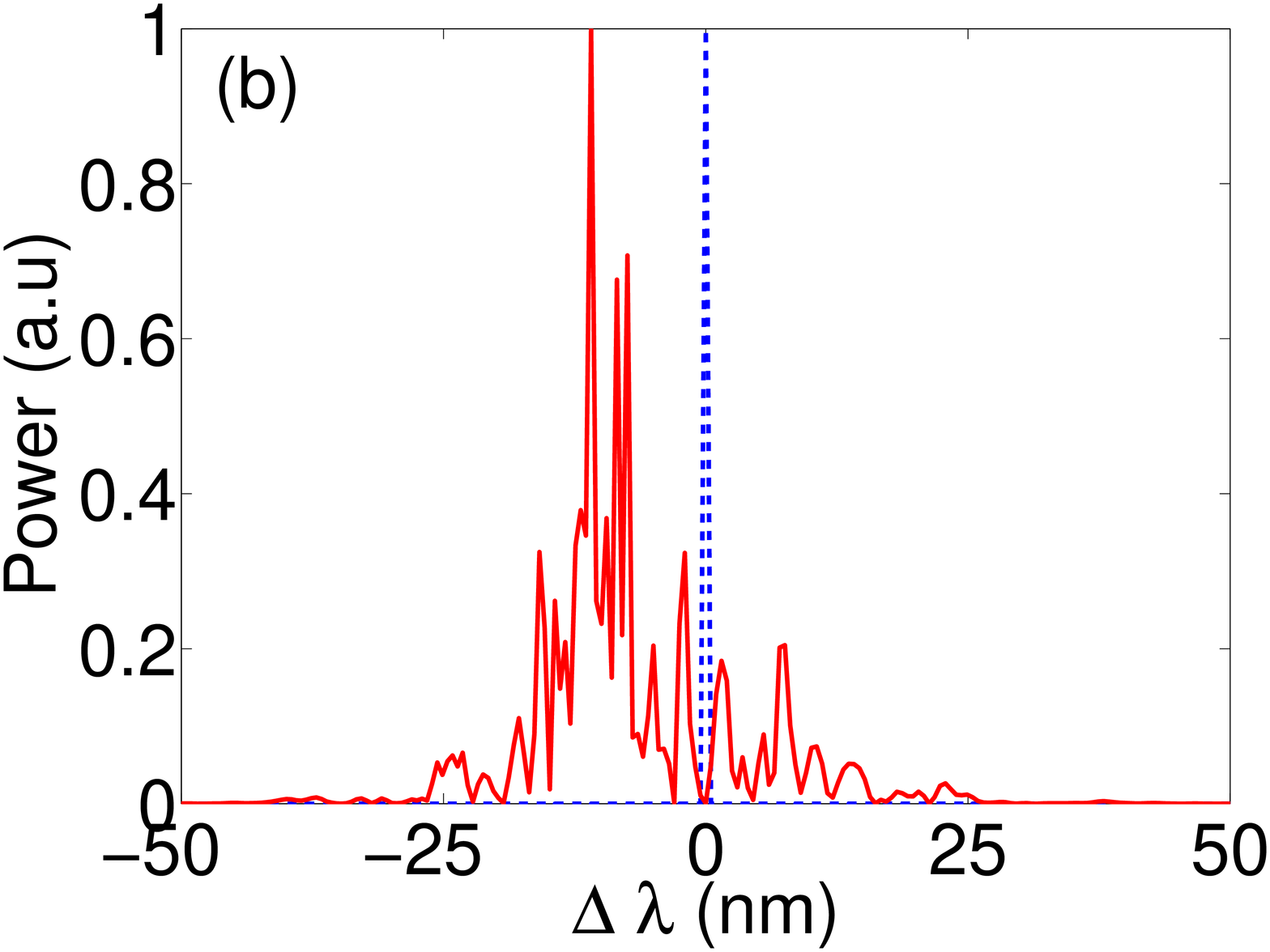}
\caption{(a) Accelerating chaotic state arising from the perturbed stationary CW nonlinear mode in AD. 
(b) Output blue-shifted spectrum (red curve) and input spectrum (dashed blue curve). In the numerical simulation we used the 
parameters  $g$=1, $g_2$=0.16, $|\beta_2|=2$ ps$^2$/m, and $K \simeq 0.4$.}
\label{Fig4}
\end{center}
\end{figure}

In order to confirm our theoretical predictions we have numerically solved Eqs. (\ref{CGLE},\ref{FCE}) by using split-step 
fast Fourier transform and Runge Kutta algorithms. Owing to the instabilities of the vacuum and of the stationary CW modes, 
noise is amplified and chaotically generates unstable accelerating pulses (see Fig. \ref{Fig4}a). Note that the intensity
of the pulses is generally higher for short times and smaller for longer times. This general behavior follows from the
influence of FCA, which is initially zero at $t \rightarrow - \infty$ and grows as time increases. Generation of FCs affects 
the pulse dynamics also via FCD, which is the sole crucial term responsible for the acceleration towards shorter times.
MI continuously generates unstable dissipative solitons that play the role of the strange attractor of the chaotic system, 
bifurcating, collapsing and creating other pulses due to their inherent instability. The characteristic duration of the 
chaotically generated pulses can be analytically predicted as $\Delta \tau \simeq 2 \pi |\Delta \Omega|^{-1}$, where  
$\Delta \Omega$ is the instability frequency window that can be calculated directly from Eq. (\ref{hE}). For $g=1$ and 
$g_2=0.16$, the analytical prediction for the pulse width is $\Delta \tau \sim$ 1.5 ps, which finds agreement with the 
numerical simulation (see Fig. \ref{Fig4}a). The temporal acceleration of the chaotically generated pulses is accompanied 
in the spectral domain by a blue-shift of about $10$ nm, as shown in Fig. \ref{Fig4}b. The main obstacle hampering 
further blue-shifting is represented by the finite amplifying window of the gaining material. Thus, if other 
gaining media with larger spectral window are used, a larger blue-shift can be achieved.

\section{Conclusions}

In conclusion, in this work we have investigated analytically and numerically the propagation dynamics of continuous waves 
in an amplifying silicon-based waveguide. We modeled optical propagation using a Ginzburg-Landau equation for the optical field 
coupled with a first order differential equation accounting for the generation of free-carriers. We have derived the stationary 
nonlinear CW mode of the system and we have studied its stability against small perturbations, finding universal modulational 
instability in both anomalous and normal dispersion regimes. By numerically solving the governing equations we have observed an 
accelerating chaotic dynamics resulting from the inherent instabilities of the system. Our theoretical investigations have been 
focused on a realistically accessible setup, and our theoretical predictions can be experimentally verified.

\indent
\indent

This research was funded by the German Max Planck Society for the Advancement of Science (MPG).

\end{document}